\documentclass[aps,prc,onecolumn,superscriptaddress,showpacs,showkeys]{revtex4}
\usepackage{graphicx}
\usepackage{bm}
\usepackage{epsfig}
\usepackage{amsfonts}
\usepackage{amssymb,amscd}
\usepackage{subfigure}
\usepackage{xcolor}

\begin{document}
	
\title{$B$ - meson production at forward rapidities in $pp$ collisions at the LHC: \\
Estimating the intrinsic bottom contribution}

\author{Yuri N. {\sc Lima}}
\email{limayuri.91@gmail.com}
\affiliation{Institute of Physics and Mathematics, Federal 
  Postal Code 354,  96010-900, Pelotas, RS, Brazil}

\author{Andr\'e V. {\sc Giannini}}
\email{AndreGiannini@ufgd.edu.br}
\affiliation{
	Faculdade de Ci\^encias Exatas e Tecnologia, Universidade Federal da Grande Dourados (UFGD),
	Caixa Postal 364, CEP 79804-970 Dourados, MS, Brazil
}
\affiliation{Departamento de F\'isica, Universidade do Estado de Santa Catarina, 89219-710 Joinville, SC, Brazil}

\author{Victor P. {\sc Gon\c{c}alves}}
\email{barros@ufpel.edu.br}
\affiliation{Institute of Physics and Mathematics, Federal 
  Postal Code 354,  96010-900, Pelotas, RS, Brazil}
\affiliation{Institute of Modern Physics, Chinese Academy of Sciences,
  Lanzhou 730000, China}
	
\begin{abstract}
	The production of $B$ mesons at forward rapidities  is strongly sensitive to the behavior of the gluon and bottom distribution functions for small and large values of the Bjorken - $x$ variable. In this exploratory study, we estimate the cross - section for the $B^{\pm}$ meson production in the kinematic range probed by the LHCb detector and that will be analyzed by the future Forward Physics Facility (FPF) considering the hybrid formalism, the solution of the running coupling Balitsky - Kovchegov equation and distinct descriptions for the bottom distribution function. We assume an ansatz for the intrinsic bottom component in the proton wave function, and estimate its impact on the transverse momentum, rapidity and Feynman - $x$ distributions. Our results indicate that the presence of an intrinsic bottom strongly modifies the magnitude of the cross - section at ultra - forward rapidities ($y \ge 6$), which implies an enhancement of the $B^{\pm}$ production at the FPF. Possible implications on the prompt neutrino flux at ultra-high energies are also briefly discussed.
\end{abstract}
	
\maketitle
\date{\today}
	
	
The production of heavy mesons in $pp$ collisions at the Large Hadron Collider (LHC) is one of the main tests of the perturbative Quantum Chromodynamics (pQCD) as well an important probe of the partonic structure of the proton at high energies \cite{Andronic:2015wma}. In particular, the study of its production at forward rapidities, probes projectile partons with large light cone momentum fractions ($x_1 \rightarrow 1$) and target partons carrying a very small momentum fraction
($x_2 \ll 1$). Consequently, it is expected to provide important constraints on the small-$x$ effects expected to be present in the target coming from the
non-linear aspects of QCD \cite{Gelis:2010nm},  and on the large-$x$ effects in the projectile, as e.g. an intrinsic heavy quark component on the proton wave function \cite{Brodsky:2020zdq}. Such expectations have motivated several phenomenological studies in recent years, mainly focused on the $D$ - meson production ~\cite{Goncalves:2008sw,LYONNET2015,Bailas:2015jlc,Boettcher:2015sqn,Carvalho:2017zge,Giannini:2018utr,Abdolmaleki:2019tbb,Maciula:2020dxv,Goncalves:2021yvw,Maciula:2022lzk,Ball:2022qks,Gauld:2023zlv,Lima:2024ksd}, considering different approaches to treat the heavy quark production and distinct models for the intrinsic charm (IC) component of the proton based on the global analysis performed in Refs. \cite{Pumplin:2007wg,Dulat:2013hea,NNPDF:2023tyk,Guzzi2023}. These analyses indicated that the current experimental data can be described by distinct theoretical approaches for the QCD dynamics and cross - section factorization, and that an intrinsic charm component has important implications on the associated $Z + D$ production at the LHCb \cite{LHCb:2021stx} and in the predictions of the $D$ - meson production at ultra - forward rapidities, which will be probed by the future Forward Physics Facility (FPF) \cite{Anchordoqui:2021ghd,Feng:2022inv} and that determines the contribution of the prompt atmospheric neutrino flux measured at the IceCube \cite{Goncalves:2017lvq}. 

\begin{figure}[t]
\begin{tabular}{ccc}
\includegraphics[scale=0.27]{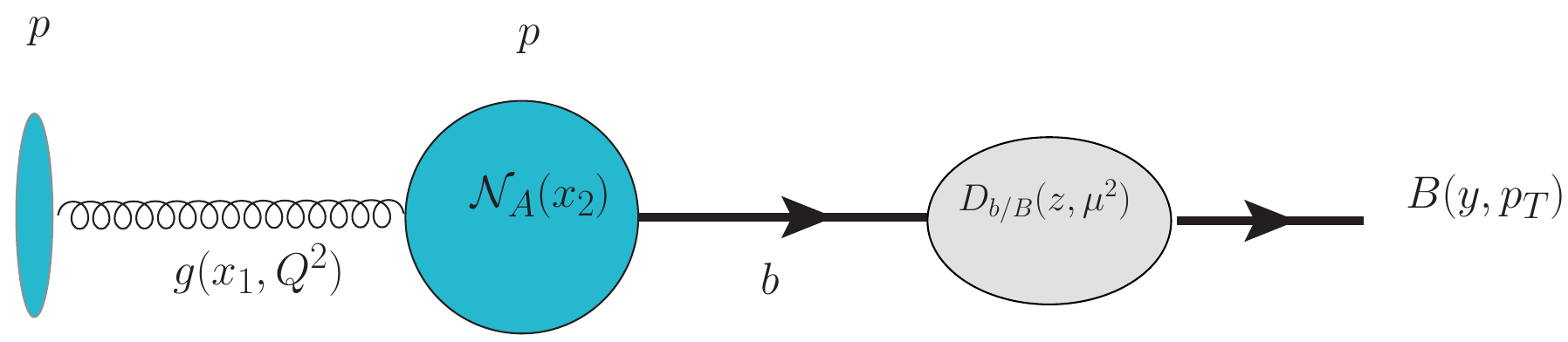} & \, & \includegraphics[scale=0.27]{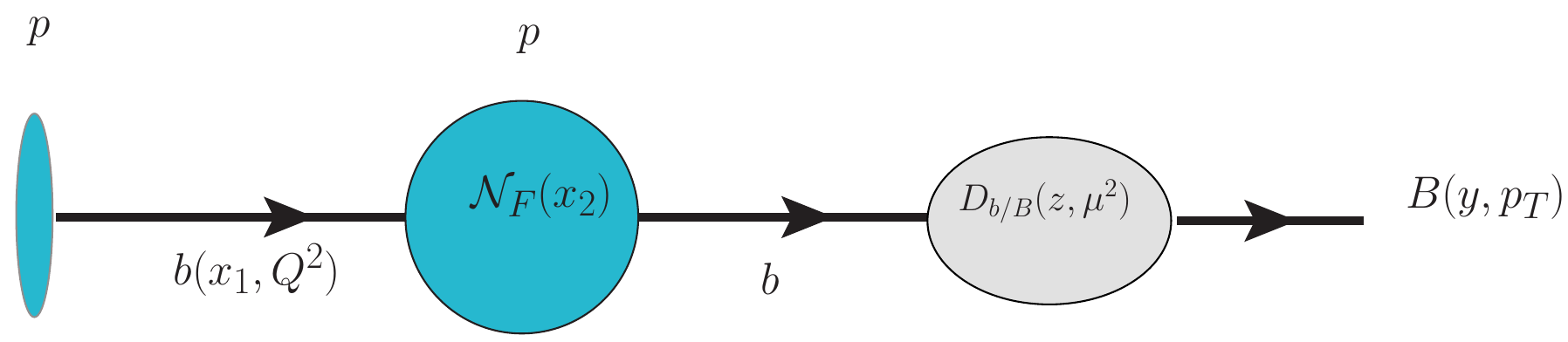}
\end{tabular}
\caption{Gluon (left)  and bottom  - initiated (right) contributions for the  production of a $B$ meson with transverse momentum $p_T$ and rapidity $y$ in $pp$ collisions.}
\label{Fig:diagram1}
\end{figure}

Over the last years, in addition to the experimental analysis of the  $D$ - meson production, the LHCb Collaboration has also measured the $B$ - meson  cross-section at forward rapidities and provided experimental data for the transverse momentum and rapidity distributions \footnote{For recent theoretical studies of the $B$ - meson production using the collinear factorization formalism see, e.g., Refs. \cite{Catani:2020kkl,Helenius:2023wkn,Mazzitelli:2023znt}.}. Two basic questions that arise are: (a) If an intrinsic bottom component is present in the proton wave function, what is its  impact  on the predictions  for the $B$ - meson production?    And (b) Are the theoretical approaches that provide a satisfactory description  of the $D$ - meson  production also able to describe the $B$ meson data? Our goal in this paper is to provide answers for these questions. In particular, the $B$ - meson production cross - section will be estimated using the hybrid formalism, which is based on the Color Glass Condensate (CGC) framework \cite{CGC}, and that used to describe the $D$ - meson production in Refs. \cite{Carvalho:2017zge,Giannini:2018utr,Bhattacharya:2016jce,Tuchin:2004rb,Goncalves:2006ch,Cazaroto:2011qq,Fujii:2013yja,
Altinoluk:2015vax,Goncalves:2017chx,SampaiodosSantos:2021tfh}. We will take  into account of gluon and bottom - initiated contributions, which are represented schematically as follows (See Fig. \ref{Fig:diagram1})
\begin{eqnarray}
\sigma(pp \rightarrow B X) \propto  g(x_1,Q^2) \otimes {\cal{N}_A}(x_2) \otimes D_{b/B} + b(x_1,Q^2) \otimes {\cal{N}_F}(x_2) \otimes D_{b/B} \,\,,
\label{Eq:sig_imp}
\end{eqnarray}
where $g(x_1,Q^2)$ and $b(x_1,Q^2)$ are the gluon and bottom densities of the projectile proton, $D_{b/B}$ is the bottom fragmentation function into a $B$ meson and the functions ${\cal{N}_A}(x_2)$ and ${\cal{N}_F}(x_2)$ are the adjoint and fundamental forward scattering amplitudes, which encodes all the information about the hadronic scattering, and thus about the non-linear and quantum effects in the proton wave function (see below). The gluon initiated contribution  can be associated to the $gg \rightarrow b \bar{b}$ subprocess, and is expected to dominate at high center - of - mass energies and central rapidities. On the other hand, the contribution of the bottom - initiated subprocess is strongly dependent on the amount of bottom quarks in the proton wave function. In general, the global analysis performed to determine the parton distribution functions (PDFs), assume that the bottom quark is not present at the initial scale of the DGLAP evolution, and that they are generated perturbatively by gluon splitting during the evolution for larger hard scales. Such a bottom content is usually denoted as extrinsic. In contrast, an intrinsic component can also be present \cite{Brodsky:1980pb}, associated with  e.g.  an $|uud b\bar{b}\rangle$ Fock component of the proton's wave function, which has multiple connections to the proton's valence quarks and is sensitive to the non-perturbative structure of the proton.
Although such an intrinsic component is expected to be suppressed by a factor $m_c^2/m_b^2$ in comparison with the intrinsic charm one, the recent evidences of an intrinsic charm (IC) component \cite{LHCb:2021stx,Ball:2022qks,NNPDF:2023tyk} imply that its contribution for the $B$ - meson production could not be fully negligible. Such is one of the main motivations for the analysis that we will perform in this paper. A current shortcoming is that, different from the IC case, an intrinsic bottom component has still not be considered in the global analysis. In order to surpass this limitation, we will consider in our exploratory study the phenomenological ansatz recently proposed in Ref. \cite{Boettcher:2024yax}, which allow us to estimate the intrinsic bottom component in terms of the IC one, which has been constrained by the current data.  Such an ansatz allow us to provide 
an estimate of the  IB contribution for the $B$ - meson production in $pp$ collisions. In what follows, we will calculate the $B$ - meson production cross - section considering a solution of  the running coupling Balitsky - Kovchegov (BK) equation \cite{BAL,KOVCHEGOV} for ${\cal{N}_A}(x_2)$ and ${\cal{N}_F}(x_2)$ in Eq. (\ref{Eq:sig_imp}), and different parameterizations, with and without an intrinsic bottom component, for the gluon and bottom PDFs. We will estimate the transverse momentum, rapidity and Feynman - $x$ distributions and a comparison with the current LHCb data \cite{LHCb20137and13TeV} will be performed. { It is important to emphasize that previous discussions about the impact of the intrinsic bottom production on the $B$ - meson production have been performed, e.g. in Refs. \cite{Vogt1995,Braaten:2001bf} using distinct theoretical approaches, mainly focused on lower center - of  - mass energies and/or central rapidities and derived disregarding non-linear effects in the QCD dynamics. As a consequence, the current study is the first analysis that takes into account of these effects and can be considered as complementary to these previous.}

Initially, let's present a brief review of the formalism used to estimate the gluon and quark - initiated contributions present in Eq. (\ref{Eq:sig_imp}), which have been discussed in detail in Refs.~\cite{Carvalho:2017zge,Goncalves:2017chx}. At forward rapidities and in the dipole frame, the gluon - initiated (G.I.) contribution,  represented in the left panel of Fig. \ref{Fig:diagram1},  can be factorized in terms of the projectile gluon distribution $x_1g(x_1,Q^2)$, the fragmentation function $D_{b/B}$ and the bottom quark production via the gluon - proton scattering, which is described by the cross-section for the $g + p \rightarrow b \bar{b} X$ process.  As demonstrated in Ref.~\cite{Goncalves:2017chx}, the differential distribution for the production of a $B$ meson with transverse momentum $p_T$ at rapidity $y$ can be expressed as follows:
\begin{eqnarray} \label{mes-CS}
	\left.\frac{d\sigma_{pp \rightarrow B X}}{dy d^2p_T}\right|_{G.I.} = \int_{z_{\rm min}}^1 \frac{dz}{z^2} \,
	 x_1g(x_1,Q^2) \,\int_{\alpha_{\rm min}}^1 d\alpha \frac{d^3\sigma_{gp \to b\bar{b}X}}{d\alpha d^2 q_T }\,D_{b/B} (z,\mu^2) \,,
	 \label{eq:ppMX}
\end{eqnarray}
where $z$ is the fractional light-cone momentum of the bottom $b$ carried by the meson, $\vec{q}_T = \vec{p}_T/z$,     $\alpha$ is the momentum fraction of the gluon carried by the bottom   and  $D_{b/B}$ is the fragmentation function. Moreover, one has that
\begin{eqnarray} \label{Eq:limits}
z_{\rm min}=\frac{\sqrt{m_B^2+p_T^2}}{\sqrt{s}}\,e^{y}\,\,\,\,\mbox{and} \,\,\,\,
\alpha_{\rm min} = \frac{z_{\rm min}}{z}\sqrt{\frac{m_b^2 z^2 + p_T^2}{m_B^2 + p_T^2}}\,.
\end{eqnarray}
 The differential cross-section for the $g + p \rightarrow b \bar{b} X$ process is given by~\cite{Goncalves:2017chx}
\begin{eqnarray}
\frac{d^3\sigma_{gp \to b\bar{b}X}}{d\alpha d^2 q_T } &=& 
\frac{1}{ 6 \pi}  \int \frac{d^2 \kappa_T}{\kappa_T^4}  \alpha_s(\mu^2)\, {\cal K}_{\rm dip}(x_2,\kappa_T^2)\,
\Big\{\Big[\frac{9}{8}{\cal{H}}_0(\alpha,\bar{\alpha},\vec{q}_T) - \frac{9}{4} {\cal{H}}_1(\alpha,\bar{\alpha},\vec{q}_T,\vec{\kappa}_T) \nonumber\\ 
&+& {\cal{H}}_2(\alpha,\bar{\alpha},\vec{q}_T,\vec{\kappa}_T) + \frac{1}{8}{\cal{H}}_3(\alpha,\bar{\alpha},\vec{q}_T,\vec{\kappa}_T)\Big] + 
\left[ \alpha \longleftrightarrow \bar{\alpha}\right]\Big\} \,,
\end{eqnarray}
with $\bar\alpha = 1 - \alpha$, and the auxiliary functions ${\cal{H}}_i$ defined by
\begin{eqnarray} \nonumber
{\cal{H}}_0(\alpha,\bar{\alpha},\vec{q}_T) & = & \frac{m_b^2 + (\alpha^2 + \bar \alpha^2)q_T^2}{(q_T^2 + m_b^2)^2} \,\,,\\
{\cal{H}}_1(\alpha,\bar{\alpha},\vec{q}_T,\vec{\kappa}_T)& = & \frac{m_b^2 + (\alpha^2 + \bar \alpha^2) \vec{q_T}\cdot 
(\vec{q_T} - \alpha \vec{\kappa}_T ) }{[(\vec{q_T} - \alpha \vec{\kappa}_T)^2 + m_b^2](q_T^2 + m_b^2)}
\,\,, \nonumber \\
{\cal{H}}_2(\alpha,\bar{\alpha},\vec{q}_T,\vec{\kappa}_T)& = & \frac{m_b^2 + (\alpha^2 + \bar \alpha^2) 
(\vec{q_T} - \alpha \vec{\kappa}_T )^2 }{[(\vec{q_T} - \alpha \vec{\kappa}_T)^2 + m_b^2]^2}
\,\,, \nonumber \\
{\cal{H}}_3(\alpha,\bar{\alpha},\vec{q}_T,\vec{\kappa}_T)& = & \frac{m_b^2 + (\alpha^2 + \bar \alpha^2) 
(\vec{q_T} + \alpha \vec{\kappa}_T )\cdot (\vec{q_T} - \bar{\alpha} \vec{\kappa}_T ) }
{[(\vec{q_T} + \alpha \vec{\kappa}_T)^2 + m_b^2][(\vec{q_T} - \bar{\alpha} \vec{\kappa}_T)^2 + m_b^2]}\,\,.
\end{eqnarray}
We will assume that $\mu^2$  and the hard scale $Q^2$ are equal to the square of the invariant mass of the $b\bar{b}$ pair ($M_{b\bar{b}} = 2\sqrt{m_b^2 + q_T^2}$). In addition, one has that $x_{1,2} = (M_{b\bar{b}}/\sqrt{s})\,e^{\pm y}$. { The  dipole transverse momentum distribution (TMD)   ${\cal K}_{\rm dip}$ can be expressed in terms of the Fourier transform of the adjoint forward scattering amplitude ${\cal{N}_A}$, as follows
\begin{eqnarray}
{\cal K}_{\rm dip}(x,\kappa_T^2) = \frac{C_F}{(2\pi)^3} \int d^2\vec{r} e^{- i \vec{\kappa}_T \cdot \vec{r} } \,\,  \nabla_r^2 {\cal{N}_A}(x,r) \,\,,
\end{eqnarray}
and, consequently, is determined by the QCD dynamics at high energies. In the large - $N_c$ limit the adjoint scattering amplitude  can be obtained from the fundamental dipole scattering amplitude  ${\cal{N}_F}$, which is the solution of the rcBK equation, using the following relation: $ {\cal{N}_A}(x,r) = 2{\cal{N}_F}(x,r) - {\cal{N}}^2_F(x,r)$. }
 In our analysis, we will estimate ${\cal K}_{\rm dip}$ in terms of a solution of the running coupling BK equation (see below). For the bottom-initiated (B.I.) contribution, represented in the right panel of Fig. \ref{Fig:diagram1}, we will consider the approach proposed in Ref.~\cite{Goncalves:2008sw} and rederived in the hybrid formalism in Ref.~\cite{Altinoluk:2015vax}. The basic idea is that a bottom quark, present in the initial state, scatters off with the color background field of the target proton and then fragments into a $B$-meson. The  differential cross-section for the production of $B$-meson with transverse momentum $p_T$ at rapidity $y$ is given by~\cite{Goncalves:2008sw}
\begin{eqnarray}
\left.\frac{d\sigma_{pp \rightarrow BX}}{dy d^2p_T}\right|_{B.I.} = \frac{\sigma_0}{2(2\pi)^2}\int_{x_F}^{1}dx_1\frac{x_1}{x_F}\bigg[b(x_1,Q^2)\,\tilde{{\mathcal{N}}}_F\left(\frac{x_1}{x_F}p_T,x_2\right)D_{b/B}\left(z = \frac{x_F}{x_1},\mu^2\right)\bigg]\,\,,
\label{Eq:charm}	
\end{eqnarray}
where  $x_{1,2}$ represent the momentum fraction of the projectile and target parton that interact in the scattering process and $x_F$ is the Feynman-$x$ of the produced meson, which are defined by
\begin{equation}
x_{1,2}=\frac{p_T}{z\sqrt{s}}e^{\pm y}
\qquad\qquad \mbox{and} \qquad\qquad
x_F = x_1 - x_2 \,\,.
\end{equation}
Moreover, $\tilde{{\mathcal{N}}}_F$ is the Fourier transform of the fundamental forward scattering amplitude, determined by solving the BK equation, and $\sigma_0$ is a constant obtained by fitting the HERA data using the corresponding solution.

\begin{figure}[t]
	\centering
		\includegraphics[scale=0.9]{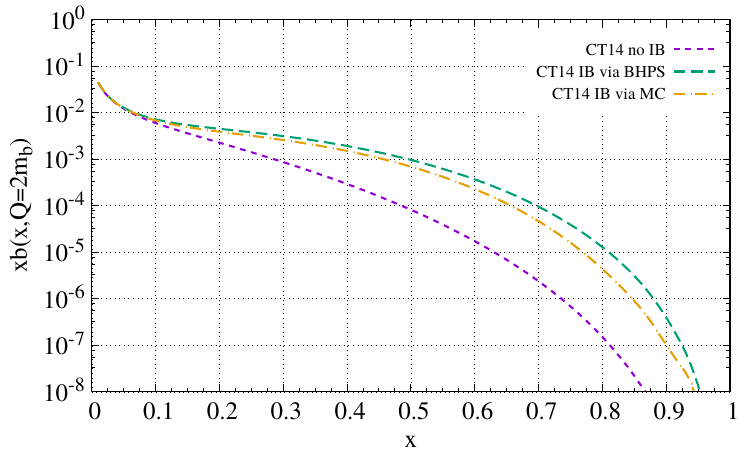}
	\caption{ Bottom distribution as a function of the Bjorken - $x$ variable for the three different PDF sets used in our study. }
	\label{Fig:pdfint}
\end{figure}

In what follows, we will present our predictions for the differential cross - sections associated with the $B^{\pm}$ production in $pp$ collisions at $\sqrt{s} = 13$ TeV.  For the adjoint and fundamental forward dipole scattering amplitudes, needed to calculate ${\cal K}_{\rm dip}$ and $\tilde{{\mathcal{N}}}_F$, we will consider the solution of the running coupling BK equation, obtained in Refs.~\cite{Albacete:2009fh,Albacete:2010sy} for an initial condition inspired by the  MV ~\cite{McLerran:1997fk} model.  The free parameters present in the initial condition are the initial saturation scale ($Q_{s,0}$) and the anomalous dimension  ($\gamma$), which are  determined by fitting the HERA data. In particular, we will consider the solution obtained for $Q_0^2 = 0.1597$ GeV$^2$ and $\gamma = 1.118$, which will denoted by ``g1.118 (MV)'' hereafter. 
For the fragmentation function $D_{b/B^{\pm}}$, we consider the BKKSS parametrization obtained in Ref. \cite{Benzke:2019usl}. However, it is important to emphasize that we have verified that similar results are obtained using other choices for the initial condition of BK equation and/or the FF parametrization obtained in Ref. \cite{Salajegheh:2019ach}. Finally, the gluon and bottom PDFs for the case without an IB component will be described by the CT14  parametrization \cite{Dulat:2013hea}. In order to estimate the impact of an intrinsic bottom component, we will consider the ansatz proposed in Ref. \cite{Boettcher:2024yax} and assume that the bottom PDF, including the extrinsic and intrinsic components, can be expressed as follows:
\begin{eqnarray}
b(x,Q^2)  = b(x,Q^2) [\mbox{no IC}] +  \frac{m_c^2}{m_b^2} \times \left\{ c(x,Q^2) [\mbox{IC}] - c(x,Q^2) [\mbox{no IC}]\right\} \,\,.
\end{eqnarray}
This ansatz takes into account the theoretical expectation that  the amount of the IB component is suppressed by a factor ${m_c^2}/{m_b^2}$ in comparison to the intrinsic charm one, which is estimated by the difference between the charm PDFs, with and without the inclusion of the IC component. Moreover, it assumes that the IB and IC components have a similar $x$ - dependence. Surely these assumptions can be improved in the future, but we believe that they provide a reasonable 
approximation for the possible amount of an intrinsic bottom in the proton wave function. In our calculations,  $b(x,Q^2) [\mbox{no IC}]$ will be parametrized by the CT14 parametrization \cite{Dulat:2013hea}. For the charm PDF, with and without IC, we will use the CTEQ6.5 parametrization, since it allow us to estimate the amount of IC for two distinct models: the Brodsky - Hoyer - Peterson - Sakai (BHPS) \cite{Brodsky:1980pb} and Meson Cloud (MC) \cite{Navarra:1995rq,Paiva:1996dd,Hobbs:2013bia} models. In the BHPS model, this component is associated to higher Fock states, as e.g. the $|uudc\bar{c}\rangle$ state. In contrast, in the MC model, it comes  e.g. from the nucleon fluctuation into an intermediate state composed by a charmed baryon plus a charmed meson.
In Fig. \ref{Fig:pdfint} we present a comparison between the associated bottom PDFs for a fixed scale $Q = 2m_b$. One has that the inclusion of an intrinsic component implies  a model dependent  enhancement of the bottom distribution at large $x$ $(> 0.1)$ by almost one order of magnitude  in comparison to the no IB prediction. As a consequence, the presence of an IB component is expected to have a large impact on the bottom initiated contribution for the $B^{\pm}$ production.

\begin{figure}[t]
\centering
\includegraphics[scale=0.7]{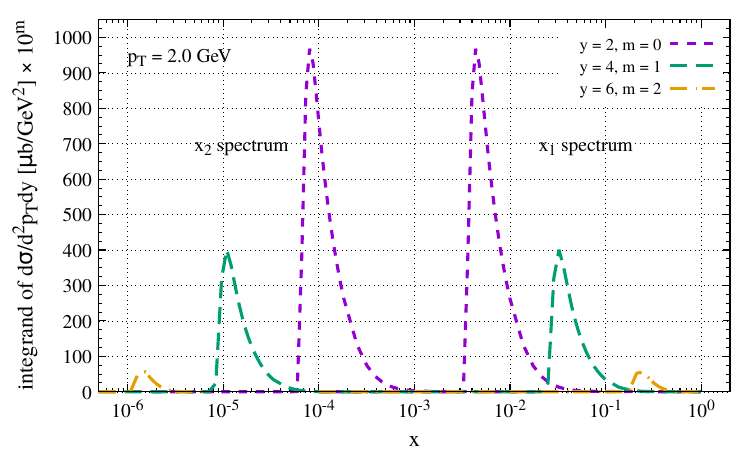}
\includegraphics[scale=0.7]{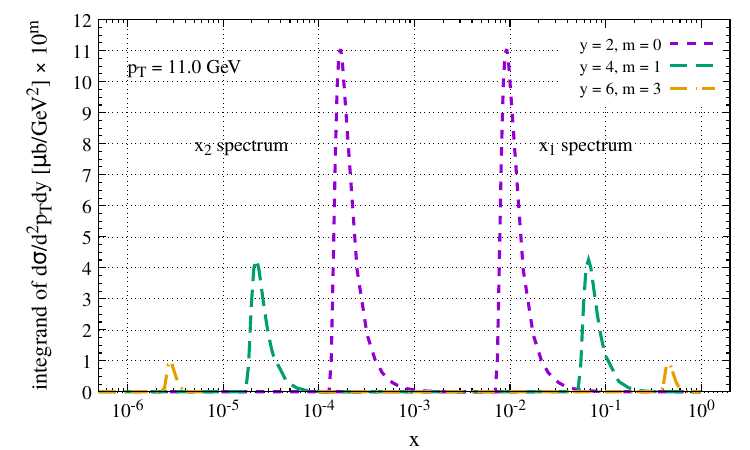}
\caption{The momentum fractions of projectile ($x_1$) and target ($x_2$) partons contributing to the differential cross-section for the $B$ - meson production, derived considering different values of the rapidity and transverse momentum of the meson. Results derived for $pp$ collisions at $\sqrt{s} = 13$ TeV.}
\label{Fig:integrand}
\end{figure}

{ Before  presenting our predictions for the differential cross-sections, it is illustrative to estimate  the typical momentum fractions of projectile ($x_1$) and target ($x_2$) partons contributing to the $B$ - meson production. In Fig.    \ref{Fig:integrand}  we present the associated results for the integrand of the differential cross-section,  derived considering $pp$ collisions at $\sqrt{s} = 13$ TeV and assuming the ``CT14 no IB" parameterization for the PDFs and the ``UGD g1.118MV" for the BK solution.  We assume different values for the rapidity and two values for the transverse momentum: $p_T = 2.0$ GeV (left panel) and  $p_T = 11.0$ GeV (right panel). The results demonstrate that for the LHC energy, the main contribution for the differential cross-section comes from values of $x_2$ smaller than 10$^{-2}$ and, therefore, is sensitive to non-linear effects in the QCD dynamics. On the other hand, one has that the contribution associated with values of $x_1$ larger than 0.1 become important for very forward rapidities, beyond the region covered by the current LHC detectors. As a consequence, considering the results presented in Fig. \ref{Fig:pdfint}, we expect a small impact of the intrinsic bottom on our predictions for the $B$ production at the LHCb rapidities ($ 2 \le y \le 4.5$). In contrast, for the future Forward Physics Facility  \cite{Anchordoqui:2021ghd,Feng:2022inv}, which will cover rapidities in the range  $6.0  \le y \le 9.0$, we can expect that the predictions will also be sensitive to the description of the bottom content of the projectile proton.}

\begin{figure}[t]
\centering
\includegraphics[scale=0.7]{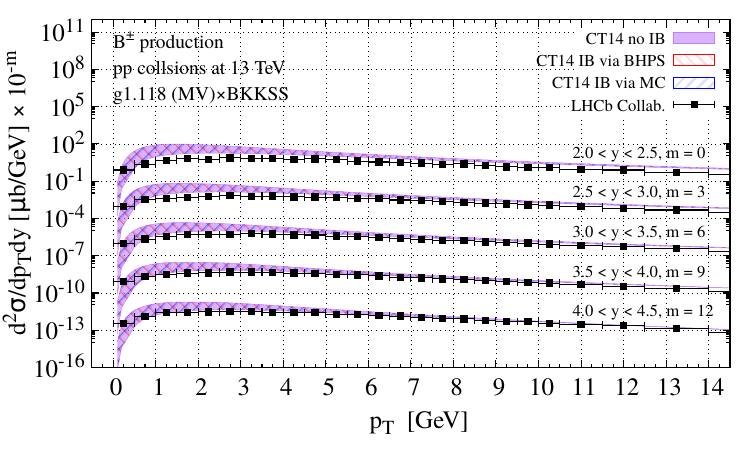}
\includegraphics[scale=0.7]{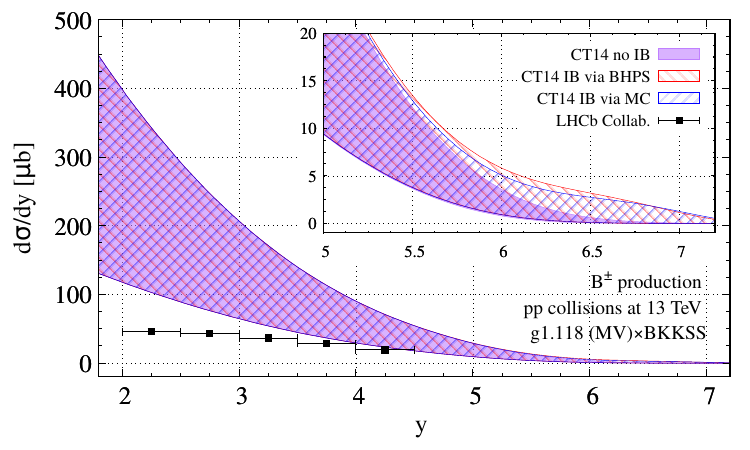}
\caption{ Transverse momentum spectra (left panel) and rapidity distribution (right panel) associated with the $B^{\pm}$ meson production in $pp$ collisions at $\sqrt{s} = 13$ TeV. Results derived using the g1.118 (MV) solution of the  rcBK equation and the BKKSS parametrization for the fragmentation function, considering  three different PDF sets. Data from  Ref.~\cite{LHCb20137and13TeV}.}
\label{Fig:pTspectrum13TeV}
\end{figure}

In Fig. \ref{Fig:pTspectrum13TeV} (left panel) we present the $p_T$-spectra of $B^{\pm}$-mesons produced in proton-proton collisions at 13 TeV. Our results, calculated using the middle value of the appropriate rapidity bin, are compared to the LHCb data~\cite{LHCb20137and13TeV}. For better visualization, we multiply the results by $10^{-m}$ ($m = 0,3,6,9,12$). The uncertainty band was generated by multiplying/dividing the momentum scale present in the PDFs and FFs by a factor of two.  A good, overall agreement with experimental data is achieved, especially for larger rapidities, with the predictions derived using distinct PDFs being similar. In particular, our results indicate that the current $B^{\pm}$ data for the $p_T$ spectra measured in the kinematical range probed by the LHCb detector is not sensitive to the intrinsic bottom. One has that the error band increases at low $p_T$, which is associated with the limitations related to the description of the incoming projectile using a perturbative QCD framework. Results derived in~\cite{Goncalves:2017chx} indicate that the agreement may be improved by taking the transverse momentum of the projectile parton into account through a phenomenological model. Another possibility is the inclusion of a Sudakov factor~\cite{Watanabe:2015yca}. However, it is important to emphasize that the treatment of the $B^{\pm}$ production at small - $p_T$ is still an open question.
	
The results for the rapidity distribution 	are presented in Fig. \ref{Fig:pTspectrum13TeV} (right panel). The rather large uncertainty band can be traced back to the fact that $p_T$ integrated observable receive major contributions from soft, low momentum particles and this is exactly the regime where our calculations for the transverse momentum spectra present the largest uncertainties when varying the momentum scale in the PDFs and FFs. The reduction of the uncertainty band at forward rapidities when compared to the mid-rapidity is purely kinematical, and happens due to the decrease of the available phase space. One has that our calculations overestimate the experimental data for smaller values of rapidity  and are closer to them for $y > 3.0$. A similar behavior is also observed in the case of the $D$ - meson production, where the data for $\sqrt{s} = 7$ TeV are quite well described, while for $\sqrt{s} = 13$ TeV the predictions overestimate the data for smaller rapidities { (See, e.g., Refs. ~\cite{Fujii:2013yja,Goncalves:2017chx,Lima:2024ksd,SampaiodosSantos:2021tfh})}. An explanation for this behaviour is still an open question, which requires a more detailed theoretical analysis as well new experimental data.  As already observed in the results for the $p_T$ - spectra, the predictions with and without an IB component in the kinematical range $2 \le y \le 4.5$  are very similar, which indicates that the inclusive  $B^{\pm}$ production at the LHCb range is not sensitive to an intrinsic bottom, {  in agreement with the expectation derived  from Fig. \ref{Fig:integrand}}. A similar conclusion is also derived when we consider the inclusive $D$ - meson production. Such result indicates that, similarly to the $D$-meson case, a better alternative for the searching of an IB component  at the LHCb could be the associated production with a $Z$ - boson, which is determined at leading order by a bottom - initiated channel. In contrast, for ultra - forward  rapidities $(y > 5)$, shown in more detail in the inset of Fig. \ref{Fig:pTspectrum13TeV}, one has that the ``CT14 no IB" prediction decreases faster than the calculations including an intrinsic bottom. These results point out that the presence of an IB component implies an enhancement of  $B^{\pm}$ production in the kinematical range that will be probed in the Forward Physics Facility. { In particular, we have verified that the ratio between the central ``IB" and ``no IB" predictions increases with the rapidity and is larger than 6 in the FPF rapidity range.}

\begin{figure}[t]
\centering
\includegraphics[scale=0.7]{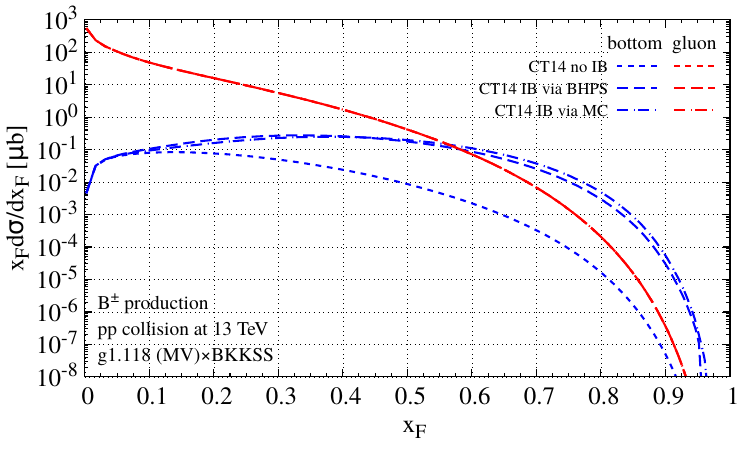}
\includegraphics[scale=0.7]{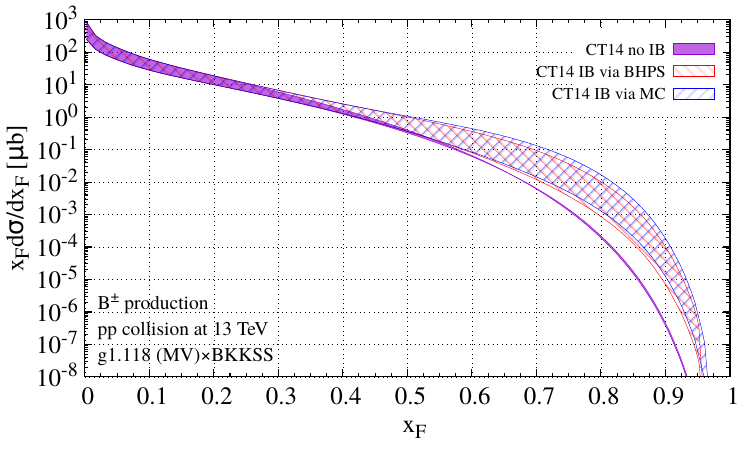}
\caption{{\bf Left panel.} Predictions of the gluon and bottom - initiated channels for the Feynman - $x$ distribution associated with the $B^{\pm}$ meson production in $pp$ collisions at $\sqrt{s} = 13$TeV.  {\bf Right panel.} Feynman-$x_F$ distributions for the sum of the gluon and bottom - initiated channels. Results derived considering three distinct PDF sets.}
\label{Fig: Feyx}
\end{figure} 

In order to determine the relative contribution of the gluon and bottom - initiated contributions for the  $B^{\pm}$ production, it is useful to analyze the corresponding  Feynman-$x_F$ distributions. In Fig. \ref{Fig: Feyx} (left panel),  we present the predictions for the $x_F$ distributions derived considering the three PDFs sets. One has that for the ``CT14 no IB" case, the  bottom - initiated channel is ever smaller than the gluon - initiated one. In contrast, the inclusion of an IB component implies that bottom - initiated channel dominates for $x_F \ge 0.6$, with the BHPS and MC predictions being similar. Such behavior has direct impact on the predictions for the $B^{\pm}$ production at large - $x_F$, as demonstrated in the right panel of Fig. \ref{Fig: Feyx}. One has that the IB component implies that the $x_F$ - distribution is enhanced by approximately one order of magnitude for $x_F \approx 0.8$. This result motivates a future analysis of the impact on the $B$ - meson production at FPF, as well a revision of the predictions for the prompt neutrino flux at the IceCube. As demonstrated in Ref. \cite{Goncalves:2017lvq}, the behavior of the Feynman - $x$ distribution at large - $x_F$ determines the magnitude of the prompt neutrino fluxes at ultra-high  energies. In addition, the results derived in Ref. \cite{Bhattacharya:2016jce} without an IB component indicate that 
the $B$ - meson contribution for the flux is $\approx 5 - 10 \%$ at $E_\nu \approx 10^5 - 10^8$ GeV, where $E_{\nu}$ is the neutrino energy. The presence of an IB component is expected to enhance this contribution, having direct impact on the predictions for the IceCube and future Neutrino Observatories. { In particular, a forthcoming study of the ratio between the tau and electron neutrino fluxes can be useful to probe the impact of the intrinsic bottom, since the electron neutrino flux at high energies is almost insensitive to the contribution associated with the $B$ - meson production, while the tau neutrino flux receives a non - negligible contribution from this channel.  }

As a summary, in this paper we have performed an exploratory study of the $B$ - meson production at forward rapidities in $pp$ collisions at $\sqrt{s} = 13$ TeV considering the hybrid formalism and taking into account 
a possible intrinsic bottom component in the proton wave function. Assuming a naive ansatz for the description of the IB component, we have demonstrated that it implies an enhancement of the bottom  PDF at large values of the Bjorken - $x$ variable. We have investigated its impact on the kinematical range probed by the LHCb detector and verified that the inclusive $B^{\pm}$ - meson production is not sensitive to the IB component. One has indicated that an alternative is the searching of intrinsic bottom at the LHCb in the $Z + B$ production. In contrast, our results demonstrated that the existence of an intrinsic bottom component implies the dominance of the bottom - initiated channel at ultra - forward rapidities, which can be probed in the  forthcoming Forward Physics Facility. Moreover, we have pointed out that the IB component will imply the enhancement of the prompt neutrino flux at high energies. Such these aspects indicate that the study of an intrinsic bottom deserves a more detailed phenomenology, which we plan to perform in forthcoming studies.

\section*{Acknowledgements}
This work was partially supported by INCT-FNA (Process No. 464898/2014-5). V.P.G. was partially supported by CNPq, CAPES and FAPERGS. Y.N.L. was partially financed by CAPES (process 001).  The authors acknowledge the National Laboratory for Scientific Computing (LNCC/MCTI, Brazil), through the ambassador program (UFGD), subproject FCNAE for providing HPC resources of the SDumont supercomputer, which have contributed to the research results reported within this study. URL: \url{http://sdumont.lncc.br}

\end{document}